\documentclass[aps,prl,twocolumn,showpacs,balancelastpage]{revtex4}

\usepackage{amsmath,amssymb,amsfonts}
\usepackage{bm}
\usepackage{epsfig}

\newcommand{\nop}[1]{#1}
\newcommand{\bra}[1]{\langle{#1}|}
\newcommand{\ket}[1]{|{#1}\rangle}
\newcommand{\braket}[2]{\langle{#1}|{#2}\rangle}

\newcommand{\gwfexp}[1]{\langle{#1}\rangle}%_{\text{G}}^{\phantom{+}}}
\newcommand{\ham}{\hat{H}}%{\hat{H}_{\text{G}}}

\begin{document}

  \title{Gossamer metals}

  \author{Marcus Kollar}

%  \email[Email address: ]{kollar@itp.uni-frankfurt.de}
  
  \affiliation{In\-sti\-tut f\"{u}r Theore\-ti\-sche Phy\-sik,
    Jo\-hann-Wolf\-gang-Goethe-Uni\-ver\-si\-t\"{a}t Frank\-furt,
    Ro\-bert-Mayer-Stra\ss{}e~8, D-60054~Frankfurt am~Main, Germany.}

  \date{August 26, 2003}

  \begin{abstract}
    Laughlin's construction of exact gossamer ground states is applied
    to normal metals. We show that for each variational parameter
    $0\leq g\leq1$, the paramagnetic or ferromagnetic Gutzwiller wave
    function is the exact ground state of an extended Hubbard model
    with correlated hopping, with arbitrary particle density,
    non-interacting dispersion, and lattice dimensionality.  The
    susceptibility and magnetization curves are obtained, showing that
    the Pauli susceptibility is enhanced by correlations.  The
    elementary quasiparticle excitations are gapless, except for a
    half-filled band at $g=0$, where a Mott transition from metal to
    insulator occurs.
  \end{abstract}


  \maketitle

  Progress in the understanding of many-body effects in strongly
  correlated electron systems, such as quantum magnets, narrow-band
  transition metal compounds, fractional quantum Hall systems, or
  high-temperature superconductors, has depended on a variety of
  theoretical tools.  Important information about the electronic
  structure can often be obtained from \emph{ab initio} calculations,
  which are however less reliable if interactions between electrons
  are dominant over their kinetic energy.  On the other hand, the
  study of idealized model systems, containing only the presumably
  relevant degrees of freedom, can provide insight into microscopic
  physical mechanisms. However, since such models are rarely exactly
  solvable, analytical and numerical calculations usually involve
  approximations or extrapolations. In view of these limitations,
  support for proposed physical notions has occasionally come from an
  inverse strategy: starting from a correlated many-body wavefunction
  one constructs a hopefully ``reasonable'' model Hamiltonian for
  which it is the exact ground state.  Correlated quantum phases may
  then be classified according to their elementary excitations or
  correlation functions. This approach has been useful in particular
  for the understanding of the fractional quantum Hall effect,
  spin-Peierls or Haldane-gap antiferromagnets, and quantum
  rotors~\cite{Arovas92}.
  
  Recently, Laughlin~\cite{Laughlin02a} developed a new approach to
  high-temperature superconductivity, viewing the insulating state as
  a superconductor with very low superfluid density.  Pursuing the
  above strategy, he proposed that the ground-state wavefunction of
  such a ``gossamer superconductor'' is obtained from the BCS
  mean-field product state by applying the Gutzwiller correlation
  operator ($0\leq g\leq1$),
  \begin{align}
    \hat{K}(g)
    =
    g^{\sum_{i}\!\hat{D}_{i}}
    =
    \prod_{\scriptstyle i}
    \left[1-(1-g)\hat{D}_{i}\right]
    ,\label{eq:K}
  \end{align}    
  where $\hat{D}_{i}=\hat{n}_{i\uparrow}\hat{n}_{i\downarrow}$ is the
  operator for double occupation at lattice site $i$, and
  constructed a corresponding model Hamiltonian.  Elementary
  excitations~\cite{Laughlin02a}, the transition from superconductor
  to Mott insulator~\cite{Zhang02a}, magnetic
  instabilities~\cite{Bernevig03a}, and related mean-field
  Hamiltonians~\cite{Bernevig03a,Yu02ff} were also studied in this
  context.
  
  The purpose of this letter is the application of Laughlin's gossamer
  paradigm to normal metals, i.e., itinerant electrons on a lattice
  without broken (discrete) translational symmetries. (In particular,
  antiferromagnetic or superconducting phases are excluded.) It is
  well known that a metallic system can be driven into an insulating
  state by strong electronic correlations.  This type of transition
  from metal to insulator, the Mott transition, occurs for example in
  transition metal oxides, and has been analyzed by a variety of
  theoretical methods \cite{Gebhard97a}.  These include the
  variational Gutzwiller wavefunction (GWF) \cite{Gutzwiller63ff}
  obtained by acting with (\ref{eq:K}) on an uncorrelated Fermi sea.
  In general the GWF describes a correlated metal, except for the
  insulating state with one immobile particle at each lattice site
  that results at $g=0$ for a half-filled band. When used as a
  variational wavefunction for the Hubbard model and evaluated within
  the Gutzwiller approximation \cite{Gutzwiller63ff}, this
  Brinkman-Rice (BR) transition \cite{Brinkman70a} occurs at finite
  critical Hubbard interaction $U_c^{\text{BR}}$.  While the
  Gutzwiller approximation becomes exact in the limit of infinite
  dimensions \cite{Metzner89a}, the BR transition is shifted to
  $U_c^{\text{BR}}=\infty$ in finite dimensions
  \cite{vanDongen89a,Gulacsi91ff}.  However, the reliability of these
  variational results is limited, as the true ground state of the
  Hubbard model in infinite dimensions may behave rather differently
  \cite{Georges96a}; for example the number of doubly occupied sites
  in general does not vanish at the transition as in the BR scenario.
  Furthermore, the analysis of elementary excitations is hampered by
  the fact that the true ground state is lower in energy, and on these
  grounds the GWF has been criticized as inadequate for describing the
  Mott transition \cite{Millis91a}.  Some of these difficulties are
  resolved for models with exact GWF ground states, which we now
  proceed to construct.

  \emph{Metallic gossamer ground state.} %
  In general a gossamer ground state is built as follows
  \cite{Laughlin02a}.  Starting from an uncorrelated product wave
  function $\ket{\phi}$ and operators
  $\hat{b}_{\bm{k}\sigma}^{\phantom{+}}$ such that
  $\hat{b}_{\bm{k}\sigma}^{\phantom{+}}\ket{\phi}=0$ for all $\bm{k}$
  and $\sigma$, one applies an invertible many-body correlator
  $\hat{K}$ to obtain a correlated wavefunction $\ket{\psi}$ $=$
  $\hat{K}\ket{\phi}$, and defines
  $\hat{\tilde{b}}_{\bm{k}\sigma}^{\phantom{+}}$ $=$
  $\hat{K}\hat{b}_{\bm{k}\sigma}^{\phantom{+}}\hat{K}^{-1}$. Then
  $\ket{\psi}$ is an exact ground state of the hermitian Hamiltonian
  $\ham$ $=$ $\sum_{\bm{k}\sigma}\tilde{E}_{\bm{k}\sigma}
  \hat{\tilde{b}}_{\bm{k}\sigma}^{+}
  \hat{\tilde{b}}_{\bm{k}\sigma}^{\phantom{+}}$
  for arbitrary $\tilde{E}_{\bm{k}\sigma}\geq0$, since $\ham\geq0$
  and $\ham\ket{\psi}=0$.
  
  In the present context we use the Gutzwiller correlator (\ref{eq:K})
  as in Refs.~\onlinecite{Laughlin02a,Zhang02a,Bernevig03a,Yu02ff},
  which is invertible for $g\neq0$, $\hat{K}(g)^{-1}=\hat{K}(g^{-1})$,
  but start from a product state containing spin-up and spin-down
  fermions, characterized by the occupation numbers
  $n_{\bm{k}\sigma}^{0}$ (with $n_{\bm{k}\sigma}^{0}$ $=$ $0$ or $1$),
  \begin{align}
    \ket{\phi}
    &=
    \!\!\!\!
    \prod_{~~~~~\bm{k}\sigma~(n_{\bm{k}\sigma}^{0}=1)~}
    \hat{c}_{\bm{k}\sigma}^{+}
    \;\ket{0}
    \,.\label{eq:phi}
  \end{align}
  This state is annihilated by the operators
  $\hat{b}_{\bm{k}\sigma}^{\phantom{+}}$ $=$
  $(1-n_{\bm{k}\sigma}^{0})\hat{c}_{\bm{k}\sigma}^{\phantom{+}}$ $+$
  $n_{\bm{k}\sigma}^{0}\hat{c}_{\bm{k}\sigma}^{+}$.  After some
  algebra, we can rewrite the Hamiltonian $\ham$ as
  \begin{align}
    \ham
    &=
    \hat{H}_t
    +
    \hat{H}_h
    +
    \hat{H}_U
    +
    \hat{H}_X
    +
    \hat{H}_Y
    +
    \text{const}
    \,,\label{eq:H_tXYU}
    \\
    \hat{H}_t
    &=
    \sum_{i\neq j,\sigma}
    \nop{T}_{ij\sigma}
    \hat{c}_{i\sigma}^{+}    
    \hat{c}_{j\sigma}^{\phantom{+}}
    \,,%\label{eq:H_t}
    ~~%\\
    \hat{H}_h
    =%&=
    -h
    \sum_{i}
    (
    \hat{n}_{i\uparrow}    
    -
    \hat{n}_{i\downarrow}    
    )
    \,,%\label{eq:H_h}
    \label{eq:H_t+h}
    \\
    \hat{H}_X
    &=
    \sum_{i\neq j,\sigma}
    X_{ij\sigma}
    (\hat{n}_{i\bar{\sigma}}
    +\hat{n}_{j\bar{\sigma}})
    \hat{c}_{i\sigma}^{+}    
    \hat{c}_{j\sigma}^{\phantom{+}}
    \,,\label{eq:H_X}
    \\
    \hat{H}_Y
    &=
    \sum_{i\neq j,\sigma}
    Y_{ij\sigma}
    \hat{n}_{i\bar{\sigma}}
    \hat{n}_{j\bar{\sigma}}
    \hat{c}_{i\sigma}^{+}    
    \hat{c}_{j\sigma}^{\phantom{+}}
    \,,%\label{eq:H_Y}
    ~~%\\
    \hat{H}_U
    =%&=
    U\sum_i
    \hat{n}_{i\uparrow}    
    \hat{n}_{i\downarrow}    
    \,,%\label{eq:H_U}
    \label{eq:H_Y+U}
  \end{align}
  with the constant term depending only on $g$ and the total particle
  density $n$ $=$ $\hat{n}_{\uparrow}+\hat{n}_{\downarrow}$, which is
  fixed; we mostly consider densities $n$ $\leq$ $1$ since
  $\ket{\psi(0)}$ $=$ $0$ otherwise.  Here $\nop{T}_{ij\sigma}$ is the
  Fourier transform of $\nop{E}_{\bm{k}\sigma}$ $=$
  $(1-2n_{\bm{k}\sigma}^0)\tilde{E}_{\bm{k}\sigma}$, and the other
  parameters are given by
  \begin{align}
    h
    &=
    -\frac{1}{2L}
    \sum_{\bm{k}\sigma}
    \sigma\,
    (1-(1+g^2)n_{\bm{k}\sigma}^0))
    \tilde{E}_{\bm{k}\sigma}
    \,,\label{eq:h}
    \\
    U
    &=
    \frac{1-g^2}{g^2L}
    \sum_{\bm{k}\sigma}
    (1-(1-g^2)n_{\bm{k}\sigma}^0))
    \tilde{E}_{\bm{k}\sigma}
    \,,\label{eq:U}
    \\
    X_{ij\sigma}
    &=
    \frac{1-g}{2g}
    \left[
    (1-g)\nop{T}_{ij\sigma}
    +
    (1+g)\tilde{T}_{ij\sigma}
    \right]
    ,\label{eq:X}
    \\
    Y_{ij\sigma}
    &=
    \frac{(1-g)^2}{2g^2}
    \left[
    (1+g^2)\nop{T}_{ij\sigma}
    +
    (1-g^2)\tilde{T}_{ij\sigma}
    \right]
    ,\label{eq:Y}
  \end{align}
  where $\tilde{T}_{ij\sigma}$ is the Fourier transform of
  $\tilde{E}_{\bm{k}\sigma}$, and $L$ is the number of lattice sites.

  A model with arbitrary non-interacting dispersion
  $\epsilon_{\bm{k}}$ can now be obtained as follows.  For given band
  dispersion $\nop{E}_{\bm{k}\sigma}$ we construct the Fermi sea via
  $n_{\bm{k}\sigma}^0$ $=$ $\Theta(-\nop{E}_{\bm{k}\sigma})$ and let
  $\tilde{E}_{\bm{k}\sigma}$ = $|\nop{E}_{\bm{k}\sigma}|$ $\geq$ $0$.
  Then we put $\nop{E}_{\bm{k}\sigma}$ $=$ $\epsilon_{\bm{k}}$ $-$
  $\epsilon_{\text{F}\!\sigma}$ and adjust the Fermi energies
  $\epsilon_{\text{F}\!\sigma}$ so that the starting wavefunction
  $\ket{\phi}$ is a Fermi sea with desired densities $n_\sigma$ $=$
  $\frac{1}{L}\sum_{\bm{k}}n_{\bm{k}\sigma}^0$.
  In the following we assume $\sum_{\bm{k}}\epsilon_{\bm{k}}$ $=$
  $0$ for convenience,
  hence $\epsilon_{0\sigma}$ $\equiv$
  $\frac{1}{L}\sum_{\bm{k}}\epsilon_{\bm{k}}n_{\bm{k}\sigma}^0$ $\leq$
  $0$.
  
  For each $0\leq g\leq1$ the Gutzwiller wavefunction $\ket{\psi(g)}$
  $=$ $\hat{K}(g)\ket{\phi}$ is the exact ground state of the extended
  Hubbard Hamiltonian (\ref{eq:H_tXYU}).  It contains the kinetic
  energy $\hat{H}_t$ of a single band $\epsilon_{\bm{k}}$, which is
  independent of $g$, and an optional Zeeman term $\hat{H}_h$, absent
  for $\hat{n}_{\uparrow}$ $=$ $\hat{n}_{\downarrow}$.  For $g<1$,
  $\ham$ contains interactions that involve at most two sites: a
  repulsive on-site interaction $\hat{H}_U$ and correlated hopping
  terms $\hat{H}_X$ and $\hat{H}_Y$, whose amplitudes are related by
  $gY_{ij\sigma}$ $=$ $(1-g)^2(\nop{T}_{ij\sigma}+X_{ij\sigma})$.
  Note that $X_{ij\sigma}$, $Y_{ij\sigma}$, and $U$ all diverge in the
  limit $g\to0$.  Similar interaction terms appear in models with
  superconducting gossamer ground states
  \cite{Laughlin02a,Zhang02a,Bernevig03a,Yu02ff}, but here those
  states cannot be lower in energy than $\ket{\psi(g)}$.  Apart from
  $g$, the magnetic field and the strength and range of the
  interactions depend on the chosen band dispersion
  $\epsilon_{\bm{k}}$ and the densities $n_\sigma$. To illustrate the
  behavior of the amplitude $\tilde{T}_{ij\sigma}$ appearing in
  (\ref{eq:X})-(\ref{eq:Y}) we now discuss several examples.

  \emph{One-dimensional systems.} %
  The dispersion for a one-dimensional ring with nearest-neighbor
  hopping $-t<0$ is $\epsilon_k=-2t\cos(k)$.
  For the Fourier transform of
  $\tilde{E}_{k\sigma}=|\epsilon_k-\epsilon_{\text{F}\!\sigma}|$ we
  find
  \begin{align}
    \tilde{T}_{j\pm1,j\sigma}
    &=
    t\,
    \big[
    2n_\sigma-1
    +
    \tfrac{1}{\pi}
    \sin(\pi n_\sigma)
    \big]
    \,,\label{eq:tildeT_1}
    \\
    \tilde{T}_{j+r,j\sigma}
    &=
    \frac{4t}{\pi(r^2-1)}
    \big[
    r\sin(\pi n_\sigma r)\cos(\pi n_\sigma)
    \\
    &~~~
    +
    \cos(\pi n_\sigma r)\sin(\pi n_\sigma)
    \big]
    \,,~~|r|\geq2
    ,\label{eq:tildeT_2}
  \end{align}
  which falls off algebraically at large distances.  At half-filling
  ($n_\sigma$ $=$ $1/2$) it is on the order of $1/r^2$ and alternates
  in sign for even $r$, while vanishing for odd $r$.  This long-range
  behavior of $\tilde{T}_{ij\sigma}$ is rather generic.  As another
  example we consider ``$1/r$'' hopping, $\nop{T}_{j+r,j\sigma}$ $=$
  $it(-1)^{r}/r$ with dispersion $\epsilon_k$ $=$ $tk$, for which the
  corresponding Hubbard model was solved by Gebhard \emph{et al.} (see
  Ref.~\onlinecite{Gebhard97a} for a review).  We obtain
  \begin{align}
    \tilde{T}_{j+r,j\sigma}
    &=
    \frac{(-1)^rt}{\pi r^2}
    \big[
    1-i\pi(2n_\sigma-1)r-e^{-2\pi in_\sigma r}
    \big]
    \,,\label{eq:tildeT_lr}
  \end{align}
  again with contributions proportional to $1/r$ (absent for
  half-filling) and $1/r^2$.
  Similar power-law behavior is typically found in dimensions $D=2,3$.

  \emph{Infinite-dimensional systems.} %
  Nearest-neigh\-bor hopping $t=1/\sqrt{2D}$ on a hypercubic lattice
  with dispersion $\epsilon_{\bm{k}}$ $=$ $-2t\sum_{\alpha=1}^D\cos
  k_\alpha$ yields the density of states $\rho_{\text{hc}}(\epsilon)$
  $=$ $\exp(-\epsilon^2/2)/\sqrt{2\pi}$ in the limit $D\to\infty$
  \cite{Metzner89a,Georges96a}.
  In order to construct the corresponding amplitude
  $\tilde{T}_{ij\sigma}$ further assumptions about its symmetry are
  necessary.  Following Ref.~\onlinecite{Bluemer03a} we assume that it
  depends only on the ``taxi-cab'' distance $||\bm{R}||$ $=$
  $\sum_{\alpha=1}^D|R_\alpha|$ and use the appropriate scaling
  $\tilde{T}_{ij\sigma}$ $=$
  $\tilde{T}_{r,\sigma}^*/\smash[b]{\sqrt{2^r\scriptstyle\binom{D}{r}}}$
  where $r$ $=$ $||\bm{R}_i-\bm{R}_j||$ $\geq$ $0$. We then obtain
  \begin{align}
    \tilde{T}_{r\sigma}^*
    &= 
    \int%\limits_{-\infty}^{\infty}
    |\epsilon-\epsilon_{\text{F}\!\sigma}|
    \,
    \rho_{\text{hc}}(\epsilon)
    \,
    \frac{\text{He}_r(\epsilon)}{\sqrt{r!}}
    \,d\epsilon
    \,,
    \\
    \tilde{T}_{2r+1,\sigma}^*
    &=
    (1-2n_\sigma)
    \,
    \delta_{r,0}
    \,,
    \\
    \tilde{T}_{2r+2,\sigma}^*
    &=
    2\,
    \rho_{\text{hc}}(\epsilon_{\text{F}\!\sigma})
    \frac{\text{He}_{2r}(\epsilon_{\text{F}\!\sigma})}{\sqrt{(2r+2)!}}
    \,,
  \end{align}
  where $\text{He}_{n}(x)$ are Hermite polynomials. At half-filling we
  find $\tilde{T}_{2r\sigma}^*\sim r^{-5/4}$, corresponding to an
  effective correlated hopping range $\sum_r\tilde{T}_{r\sigma}^*{}^2$
  of order unity.  For other densities of states $\rho(\epsilon)$, in
  particular those with finite bandwidth, it is also possible
  construct a corresponding dispersion $\epsilon_{\bm{k}}$
  \cite{Bluemer03a}, and then derive $\nop{T}_{ij\sigma}$ and
  $\tilde{T}_{ij\sigma}$ in a similar fashion.

  \emph{Response to external magnetic field.} %
  Returning to the case of arbitrary dispersion and densities, we note
  that according to the equation of state (\ref{eq:h}) the
  ground-state magnetization $m$ $=$
  $\hat{n}_{\uparrow}-\hat{n}_{\downarrow}$ is nonzero if an external
  magnetic field $h$ is present.  For the homogeneous susceptibility
  $\chi$ we obtain
  \begin{align}
    \chi(h)^{-1}
    &=
    \frac{\partial h}{\partial m}
    =
    \frac{1}{4}
    \sum_\sigma
    \frac{
      1-(1-g^2)n_{\sigma}
    }{
      \rho(\epsilon_{\text{F}\!\sigma})
    }
    \,.\label{eq:chi_h}
  \end{align}
  \begin{figure}[t]
    \centerline{\epsfig{file=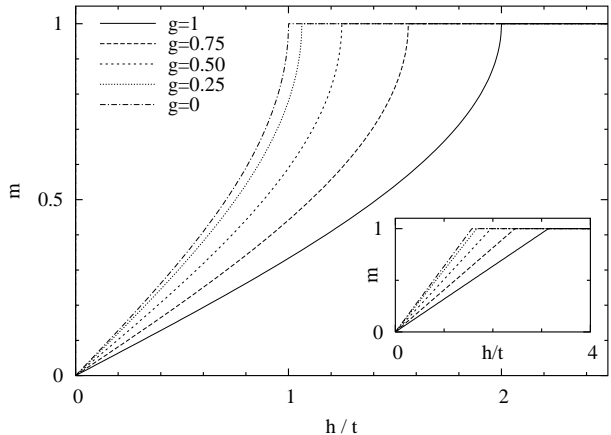,width=\hsize}}
    \caption{Magnetization curves $m(h)$
      for a one-dimensional ring with nearest-neighbor hopping at
      half-filling. Inset: same for $1/r$ hopping.
      \label{fig:m}}
  \end{figure}
  In the limit of zero field this reduces to $\chi(0)$ $=$
  $\chi_0/[1-(1-g^2)n/2]$.  As expected the system behaves like a
  correlated paramagnet, i.e., the interactions enhance the Pauli
  susceptibility $\chi_0$ $=$ $2\rho(\epsilon_{\text{F}})$ of the
  uncorrelated system.  However, it should be kept in mind that the
  interaction parameters (\ref{eq:U})-(\ref{eq:Y}) do not remain
  constant when the parameters $h$ or $m$ are varied.
  Fig.~\ref{fig:m} shows the magnetization as a function of magnetic
  field for a one-dimensional ring at half-filling. Interestingly, for
  nearest-neighbor hopping the upward curvature of these magnetization
  curves is very similar to Bethe-ansatz results for the pure Hubbard
  model \cite{Lieb68a,Takahashi69a}, where a metal-insulator
  transition occurs at $U_c=0^+$.  By contrast, for $1/r$ hopping the
  magnetization curves are strictly linear, $m$ $=$ $\chi(0)h$, due to
  the constant density of states.

  \emph{Metal-insulator transition.} %
  For an unpolarized half-filled band ($n$ $=$ $1$,
  $\epsilon_{\text{F}\!\sigma}$ $=$ $0$), the ground-state wave
  function $\ket{\psi(g)}$ describes a metal for $g>0$ and an
  insulator for $g=0$. In the insulating state there are no doubly
  occupied sites, the discontinuity of $n_{\bm{k}\sigma}$ at the Fermi
  surface vanishes, and the kinetic energy $\gwfexp{\hat{H}_t}$ is
  zero.  This Mott metal-insulator transition in the ground state of
  $\ham$ occurs at infinite interactions (\ref{eq:U})-(\ref{eq:Y}), in
  contrast to the variational BR transition, or numerical results for
  the Hubbard model in infinite dimensions \cite{Georges96a}.
  
  Nevertheless we may, somewhat artificially, shift the transition to
  finite interactions as follows. Clearly $\ket{\psi(g)}$ remains the
  ground state when we multiply $\ham$ by a positive $g$-dependent
  factor, although qualitatively different Hamiltonians may then
  result in the limit $g\to0$.  For example, for the Hamiltonian
  $\ham^{(1)} = g \ham$ the $X$ term has a finite limit, while
  $\ham^{(2)} = g^2 \ham$ yields a vanishing $X$ term and finite $Y$
  and $U$ terms; in both cases the quadratic kinetic energy vanishes
  at $g=0$.  In particular we may conclude that for any dispersion
  $\epsilon_{\bm{k}}$ the Hamiltonian
  \begin{align}
    \ham'
    &=
    \sum_{i\neq j,\sigma}
    Y_{ij}'
    \hat{n}_{i\bar{\sigma}}
    \hat{n}_{j\bar{\sigma}}
    \hat{c}_{i\sigma}^{+}    
    \hat{c}_{j\sigma}^{\phantom{+}}
    +
    U'
    \sum_i
    \hat{n}_{i\uparrow}    
    \hat{n}_{i\downarrow}    
    \,,\label{eq:H_prime}
  \end{align}
  where $Y_{ij}'$ is the Fourier transform of $\epsilon_{\bm{k}}
  (1-n_{\bm{k}\sigma}^0)$, has the exact ground state
  $\ket{\psi(g=0)}$ at half-filling if $U'$ $\geq$ $U_c'$, with
  critical interaction $U_c'$ $=$ $-\sum_\sigma \epsilon_{0\sigma}$
  $\equiv$ $|\epsilon_0|$.  Interestingly, the uncorrelated kinetic
  energy also sets the energy scale of the BR transition in the
  Gutzwiller approximation \cite{Brinkman70a}, where $U_c^{\text{BR}}$
  $=$ $8|\epsilon_0|$.  Although $\ham'$ is not a standard Hubbard
  Hamiltonian, it nonetheless appears to be the simplest model with a
  BR-type transition to an exact insulating ground state at finite
  Hubbard interaction.

  \begin{figure}[t]
    \centerline{\epsfig{file=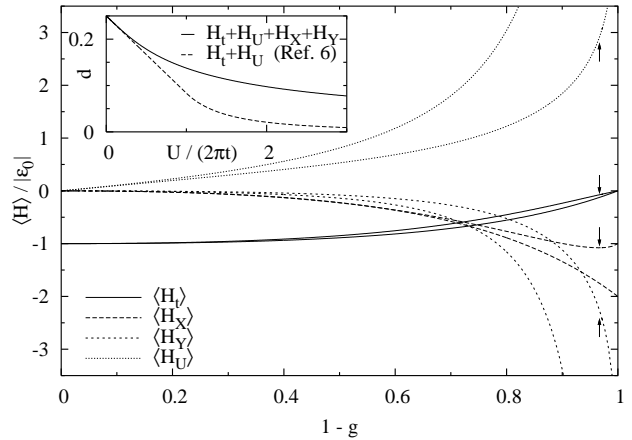,width=\hsize}}
    \caption{Expectation values of parts of $\ham$,
      (\ref{eq:H_t+h})-(\ref{eq:H_Y+U}), for nearest-neighbor hopping
      in $D$ $=$ $1$ (arrows) and $D$ $=$ $\infty$ at half-filling.
      Inset: double occupancy for $1/r$ hopping in $D$ $=$ $1$, as
      compared to the pure Hubbard model.\label{fig:H_tXYU}}
  \end{figure}

  \emph{Ground-state expectation values.} %
  The separate expectation values of the kinetic energy $\hat{H}_t$
  and the interaction terms $\hat{H}_X$, $\hat{H}_Y$, and $\hat{H}_U$
  can be calculated from the quantities $n_{\bm{k}\sigma}$ $=$
  $\gwfexp{\hat{c}_{\bm{k}\sigma}^{+}
    \hat{c}_{\bm{k}\sigma}^{\phantom{+}}}$, $d$ $=$ $\frac{1}{L}
  \sum_{i} \gwfexp{\hat{n}_{i\uparrow} \hat{n}_{i\downarrow}}$,
  $x_{ij\sigma}$ $=$ $\gwfexp{\hat{n}_{i\bar{\sigma}}
    \hat{c}_{i\sigma}^{+} \hat{c}_{j\sigma}^{\phantom{+}}}$, and
  $y_{ij\sigma}$ $=$ $\gwfexp{\hat{n}_{i\bar{\sigma}}
    \hat{n}_{j\bar{\sigma}} \hat{c}_{i\sigma}^{+}
    \hat{c}_{j\sigma}^{\phantom{+}}}$.  Using the methods of
  \cite{Metzner87ff,Kollar01ff} it can be shown that for the GWF the
  Fourier transforms of the latter are given by
  \begin{align}
    x_{\bm{k}\sigma}
    &=
    n_{\bm{k}\sigma}^{0}
    \bigg[
      n_{\bar{\sigma}}
      -\frac{1-n_{\bm{k}\sigma}}{1-g}
    \bigg]
    -
    \frac{(1-n_{\bm{k}\sigma}^{0})g\,n_{\bm{k}\sigma}}{1-g}
    -
    d
    \,,\label{eq:xks-alternate}
    \\
    y_{\bm{k}\sigma}
    &=
    n_{\bm{k}\sigma}^{0}
    \bigg[
      n_{\bar{\sigma}}
      -\frac{1-n_{\bm{k}\sigma}}{(1-g)^2}
    \bigg]
    +
    \frac{(1-n_{\bm{k}\sigma}^{0})
        g^2n_{\bm{k}\sigma}}{(1-g)^2}
    -
    d
    \,,\!\!
    \label{eq:yks-alternate}
  \end{align}
  i.e., only the momentum-space distribution $n_{\bm{k}\sigma}$ and
  the double occupancy $d$ are needed. They may be evaluated
  numerically by Monte-Carlo methods, but are also available in closed
  form under certain circumstances.  For one-dimensional systems with
  symmetric Fermi sea ($n_{k\sigma}^{0}$ $=$ $n_{-k\sigma}^{0}$) both
  quantities have been calculated analytically
  \cite{Metzner87ff,Kollar01ff}.  In dimension $D=2,3$ high-order
  perturbative methods can be used \cite{Gulacsi91ff}.  The Gutzwiller
  approximation \cite{Gutzwiller63ff}, with piecewise constant
  momentum distribution $n_{\bm{k}\sigma}$, is recovered in infinite
  dimensions \cite{Metzner89a}.
  
  The expectation values of the various parts of $\ham$ are shown in
  Fig.~\ref{fig:H_tXYU} for nearest-neighbor hopping in $D$ $=$ $1$
  and $D$ $=$ $\infty$ at half-filling. We note that
  $\gwfexp{\hat{H_X}}$ approaches a constant for $g\to0$, while
  $\gwfexp{\hat{H_Y}}$ and $\gwfexp{\hat{H_U}}$ diverge. This behavior
  occurs for all $D$, since $d\sim g^2\ln(1/g)$ in one dimension
  \cite{Metzner87ff}, $d=o(g)$ in all finite dimensions
  \cite{vanDongen89a,Gulacsi91ff}, and $d$ $\sim$ $g$ in infinite
  dimensions. We may thus conclude that the penalty that $\hat{H}_U$
  imposes on double occupancies is compensated by assisted hopping due
  to the nonstandard three-body interaction $\hat{H}_Y$.
  
  The effect of correlated hopping is also apparent when comparing to
  the pure Hubbard ring with $1/r$ hopping, which features a
  metal-insulator transition at $U_c$ $=$ $2\pi t$ with continuous
  nonzero double occupancy $d$ \cite{Gebhard97a}.  For comparison with
  previous studies of variational wavefunctions in the vicinity of
  this transition \cite{Gebhard94a,Dzierzawa95a}, $d$ vs.\ $U$ is
  shown in the inset of Fig.~\ref{fig:H_tXYU}.  The results for both
  models with $1/r$ hopping agree for weak interactions, but the
  energy gain from correlated hopping leads to a larger number of
  doubly occupied sites for strong coupling in the model
  (\ref{eq:H_tXYU}), as expected.

  \emph{Quasiparticle excitations.} %
  The known ground state of $\ham$ suggests that it might also be
  possible to calculate dynamical properties of the model, such as the
  spectral function. Unfortunately the construction of exact excited
  states is not straightforward, be it with one added or removed
  particle, or with charge or spin excitations. We therefore proceed
  by considering the variational states
  \cite{Laughlin02a,Buenemann03a}
  \begin{align}
    \ket{\bm{k}\sigma}
    =
    \hat{K}\hat{b}_{\bm{k}\sigma}^{+}\ket{\phi}
    =
    \left\{
      \begin{array}{ll}
        \hat{K}\hat{c}_{\bm{k}\sigma}^{+}\ket{\phi}
        &
        \text{~if~}n_{\bm{k}\sigma}^{0}=0
        \\
        \hat{K}\hat{c}_{\bm{k}\sigma}^{\phantom{+}}\ket{\phi}        
        &
        \text{~if~}n_{\bm{k}\sigma}^{0}=1
      \end{array}
    \right.
    \,,
  \end{align}
  whose mean energy is
  \begin{align}
    \nop{E}_{\bm{k}\sigma}^{\pm}
    =
    \frac{
      \bra{\bm{k}\sigma}
      \ham
      \ket{\bm{k}\sigma}}{\braket{\bm{k}\sigma}{\bm{k}\sigma}}
    =
    \frac{
      \braket{\psi}{\psi}}{
      \braket{\bm{k}\sigma}{\bm{k}\sigma}}
      \tilde{E}_{\bm{k}\sigma}\;
    \,,
  \end{align}
  where the commutator relations
  ${[\hat{b}_{\bm{k}\sigma}^{\phantom{+}},}$
  ${\hat{b}_{\bm{k'}\sigma'}^{+}]}$ $=$
  $\delta_{\bm{k}\bm{k}'}\delta_{\sigma\sigma'}$ and
  ${[\hat{b}_{\bm{k}\sigma}^{\phantom{+}},}$
  ${\hat{b}_{\bm{k'}\sigma'}^{\phantom{+}}]}$ $=$ $0$ were used. The
  states $\ket{\bm{k}\sigma}$ are mutually orthogonal and their energy
  is thus an upper bound to the quasiparticle energy for momentum
  $\bm{k}$ and spin $\sigma$.  The variational energy to add a
  particle (i.e., $n_{\bm{k}\sigma}^{0}=0$) is
  \begin{align}
    \nop{E}_{\bm{k}\sigma}^{+}
    &=
    \frac{
      |\epsilon_{\bm{k}}-\epsilon_{\text{F}\!\sigma}|
    }{
      1-(1+g)[(1-g)n_{\bar{\sigma}}+(1+g)n_{\bm{k}\sigma}]
    }
    \,,
  \end{align}
  while for the removal of a particle (with $n_{\bm{k}\sigma}^{0}=1$)
  \begin{align}
    \nop{E}_{\bm{k}\sigma}^{-}
    &=
    \frac{
      g^2
      |\epsilon_{\bm{k}}-\epsilon_{\text{F}\!\sigma}|
    }{
      (1+g)[(1-g)n_{\bar{\sigma}}+(1+g)n_{\bm{k}\sigma}]-1-2g
    }
    \,.
  \end{align}
  \begin{figure}[t]
    \centerline{\epsfig{file=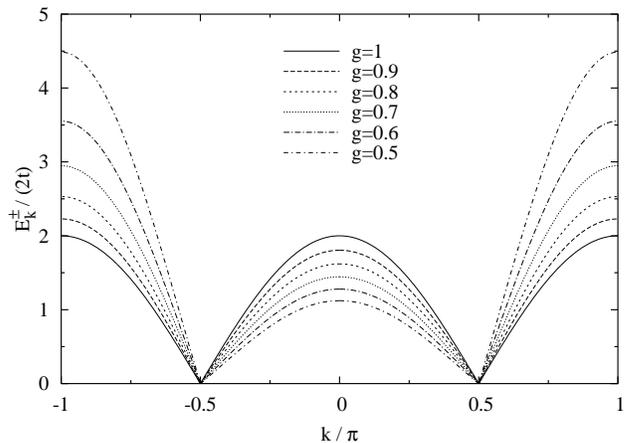,width=\hsize}}
    \caption{Quasiparticle excitations in a one-dimensional
      ring with nearest-neighbor hopping $t>0$.\label{fig:E}}
  \end{figure}
  Clearly the quasiparticle excitations are gapless, since
  $\nop{E}_{\bm{k}\sigma}^{\pm}$ $\to$ $0$ close to the Fermi surface.
  Fig.~\ref{fig:E} shows these energies for one-dimensional
  nearest-neighbor hopping at half-filling.

  \emph{Conclusion.} %
  We have constructed and characterized a new class of itinerant
  electron models for which the metallic Gutzwiller wavefunction is an
  exact ground state, due to the interplay of Hubbard interaction and
  correlated hopping.  For a half-filled band a Mott metal-insulator
  transition similar to the Brinkman-Rice scenario occurs,
  illustrating Mott's original idea of a quantum phase transition
  entirely due to charge correlations without magnetic ordering.
  Further study of the elementary excitations in these models should
  be fruitful.

  This work was supported in part by the DFG via Forschergruppe
  FOR~412.

\end{document}